\documentclass[aps,prl,reprint,groupedaddress,superscriptaddress]{revtex4-1}

\usepackage[pdftex]{graphicx}
\usepackage{amssymb,amsfonts,amsmath}
\usepackage{grffile}

\begin{document}


\title{Clustering as a prerequisite for chimera states in globally 
coupled systems}


\author{Lennart Schmidt}
\affiliation{Physik-Department, Nonequilibrium Chemical Physics, Technische Universit\"{a}t M\"{u}nchen,
  James-Franck-Str. 1, D-85748 Garching, Germany}
\affiliation{Institute for Advanced Study - Technische Universit\"{a}t M\"{u}nchen,
  Lichtenbergstr. 2a, D-85748 Garching, Germany}

\author{Katharina Krischer}
\email[]{krischer@tum.de}
\affiliation{Physik-Department, Nonequilibrium Chemical Physics, Technische Universit\"{a}t M\"{u}nchen,
  James-Franck-Str. 1, D-85748 Garching, Germany}


\date{\today}

\begin{abstract}
The coexistence of coherently and incoherently oscillating parts in a 
system of identical oscillators with symmetrical coupling, i.e. a
chimera state, is even observable with uniform global coupling. We 
address the question of the prerequisites for these states to occur in globally coupled 
systems. By analyzing two different types of 
chimera states found for nonlinear global coupling, we show that a 
clustering mechanism to split the ensemble into two groups is needed 
as a first step. In fact, the chimera states inherit properties from 
the cluster states in which they originate. Remarkably, they can exist 
in parameter space between cluster and chaotic states, as well as 
between cluster and synchronized states.
\end{abstract}


\maketitle


The story of chimeras in nonlinear dynamics goes back to the year 2002, when 
Kuramoto \& Battogtokh \cite{Kuramoto_NPCS_2002} discovered that in a 
system of identical oscillators with symmetrical coupling, coherently 
oscillating regions can coexist with incoherent ones. In fact, this
was not the first observation of such a coexistence
\cite{Umberger_PRA_1989, Kaneko_PhysicaD_1990, Nakagawa_PTP_1993}, but
Kuramoto \& Battogtokh were the first pointing out its importance. This state was 
named a chimera state by Abrams \& Strogatz 
\cite{AbramsStrogatz_PRL_2004}, referring to the chimera in greek 
mythology. Many theoretical studies followed, see for example 
Refs.~\cite{AbramsStrogatz_PRL_2004, ShimaKuramoto_PRE_2004, 
Abrams_PRL_2008, Bordyugov_PRE_2010, Laing_PRE_2010, 
Schoell_PRL_2011, Yeldesbay_PRL_2014, Sethia_PRL_2014}, and 10 years
after their discovery chimera states could be observed in experiments also 
\cite{Tinsley_Nature_2012, Hagerstrom_Nature_2012, 
Martens_PNAS_2013, Nkomo_PRL_2013, Wickramasinghe_PONE_2013, 
Schmidt_Chaos_2014}. For a recent review see 
Ref.~\cite{Panaggio_arXiv_Review}. 
Yet, concerning the prerequisites of their existence and the mechanisms
of their emergence only very little is known. Bifurcation analysis
revealed that they can emerge via a saddle-node bifurcation 
\cite{AbramsStrogatz_PRL_2004, Abrams_PRL_2008, Laing_PRE_2010, 
Laing_PhysicaD_2009}, and 
they were found in maps with coupling-induced bistability  
\cite{Schoell_PRL_2011}. First analytical studies aiming to analyze the stability and to
characterize the emergence and dynamics of chimera states in
nonlocally coupled systems in a general way are presented in 
Refs.~\cite{Laing_PhysicaD_2009, Omelchenko_Nonlinearity_2013}.
In addition, it has long been thought that a nonlocal coupling scheme is indispensible for 
their formation. Under nonlocal coupling it is reasonable that regions 
of different dynamics can coexist, since the coupling decreases with 
the distance and the influence of one region on the other over some 
interfacial region might not be 
too strong. However, it could be shown that they also exist in systems 
with uniform global coupling \cite{Schmidt_Chaos_2014, Daido_PRL_2006, 
Sethia_PRL_2014, Yeldesbay_PRL_2014}, where each oscillator is 
influenced equally strongly by all the other oscillators. 
Such a coupling is realized experimentally e.g. by an external resistance in series
with some voltage-controlled device, such as a gas-discharge tube \cite{Purwins_AP_2010} or
an electrochemical cell \cite{Krischer_2003} or due to 
rapid mixing in the gas phase in surface reactions
\cite{Mertens_JCP_1993, Mertens_JCP_1994}. Generally, it arises
whenever a global quantity is
controlled and can be linear, as well as nonlinear.
For globally coupled phase oscillators chimera states were found in a system with
time delay, where bistability emerged in a self-consistent
way, as well as with individual bistable oscillators
\cite{Yeldesbay_PRL_2014}. 

In this Letter we argue that a clustering mechanism observed typically in
globally coupled systems is a sufficient feature, rendering
chimera states possible, as it splits the
oscillators into several groups and yields at least bistability. Then, one of the two groups can
desynchronize, while the other group stays coherent if the
response on the coupling is effectively different in the two groups. 
In the present study we demonstrate that this situation can arise via nonlinear 
amplitude effects.
Moreover, we show that different cluster states lead to different chimera states 
and that the chimera states inherit properties from the cluster states 
in which they originate.

Our system is composed of $N$ Stuart-Landau oscillators, each of the form
\begin{equation}
  \frac{\mathrm d}{\mathrm dt} W_k = W_k - (1+i c_2) \left| W_k
  \right|^2 W_k \ ,
\end{equation}
$k=1,2,\dots,N$, constituting generic limit-cycle oscillators near a Hopf bifurcation 
\cite{Kuramoto_2003}. 
We couple them via a nonlinear global coupling:
\begin{align}
  \frac{\mathrm d}{\mathrm dt} W_k = &W_k - (1+i c_2) \left| W_k
  \right|^2 W_k \notag \\
  &- (1+i \nu) \left< W \right> + (1+i c_2)
  \left< \left| W \right|^2 W \right> \ .
\label{eq:SL_ensemble}
\end{align}
Here $\left< \cdots \right>$ describes the arithmetic mean of the
oscillator population, i.e. $\left< W \right> = \sum_{k=1}^N W_k /
N$. Taking the average of the whole equation yields for the dynamics
of the mean value
\begin{equation}
  \frac{\mathrm d}{\mathrm dt} \left< W \right> = - i \nu \left<
    W \right> \quad \Rightarrow \quad \left< W \right> = \eta
  e^{-i \nu t} \ .
\label{eq:conservation_law}
\end{equation}
\begin{figure*}[ht]
  \centering
  \includegraphics[width=15cm]{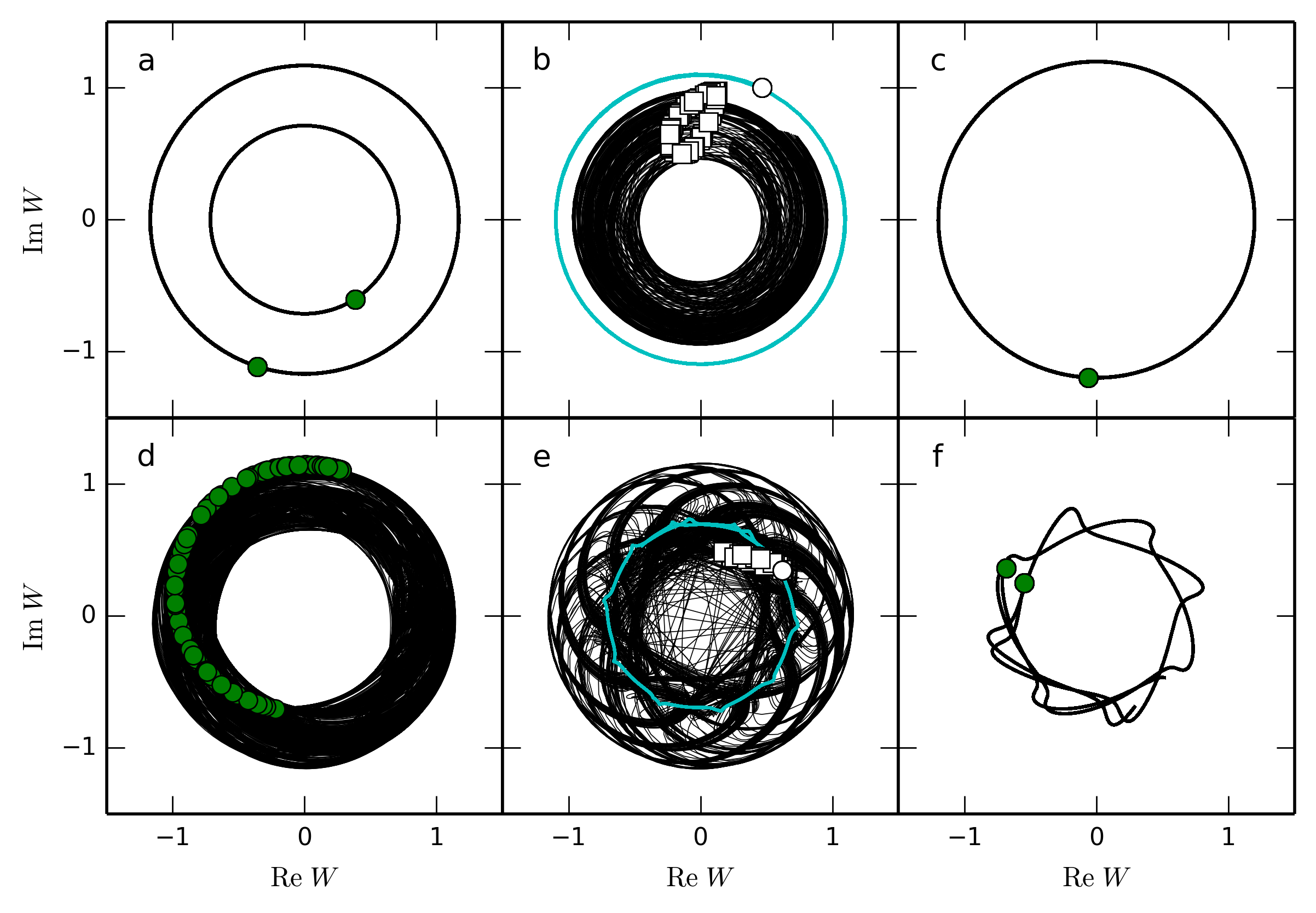}
  \caption{(Color online) Evolutions in the complex plane and
    snapshots. Trajectories of the oscillators are shown as solid
    lines, whereas the symbols describe snapshots of the system. First
    row: type I dynamics. (a) Amplitude
    clusters ($\eta = 0.9$). (b) Type I chimera ($\eta = 1.02$),
    black lines and squares: incoherent group; cyan (gray) lines and
    circles: coherent group. (c) Complete synchronization ($\eta = 1.2$). Other
    parameters: $c_2 = 0.58$ and $\nu = 1.49$. Second row: type II
    dynamics. (d) Irregular dynamics ($\nu = -0.1$). (e) Type II chimera ($\nu = 0.02$),
    black lines and squares: incoherent group; cyan (gray) lines and circles: coherent group.
    (f) Two-phase clusters ($\nu = 0.1$). Other parameters: $c_2 = -0.6$, $\eta
    = 0.7$.}
  \label{fig:complexp}
\end{figure*}
This constitutes conserved harmonic oscillations of the ensemble 
average. Note that the above model, Eq.~\eqref{eq:SL_ensemble}, 
describes the essential dynamics of the oxide-layer thickness during 
the photoelectrodissolution of n-type silicon \cite{Miethe_PRL_2009, 
GarciaMorales_PRE_2010, Schmidt_arXiv_ECCII}. The linear global coupling is a result
of an external resistance in series with the silicon
electrode, while we believe that the nonlinear global coupling is
connected to a limitation of the total amount of charge carriers. For
more details, see Refs.~\cite{Schmidt_arXiv_ECCII,
Schoenleber_NJP_2014}. This experimental system exhibits also 
chimera states \cite{Schmidt_Chaos_2014, Schoenleber_NJP_2014}. To 
capture its dynamics it is important to reproduce the harmonic 
mean-field oscillation.

\begin{figure*}[ht]
  \centering
  \includegraphics[width=15cm]{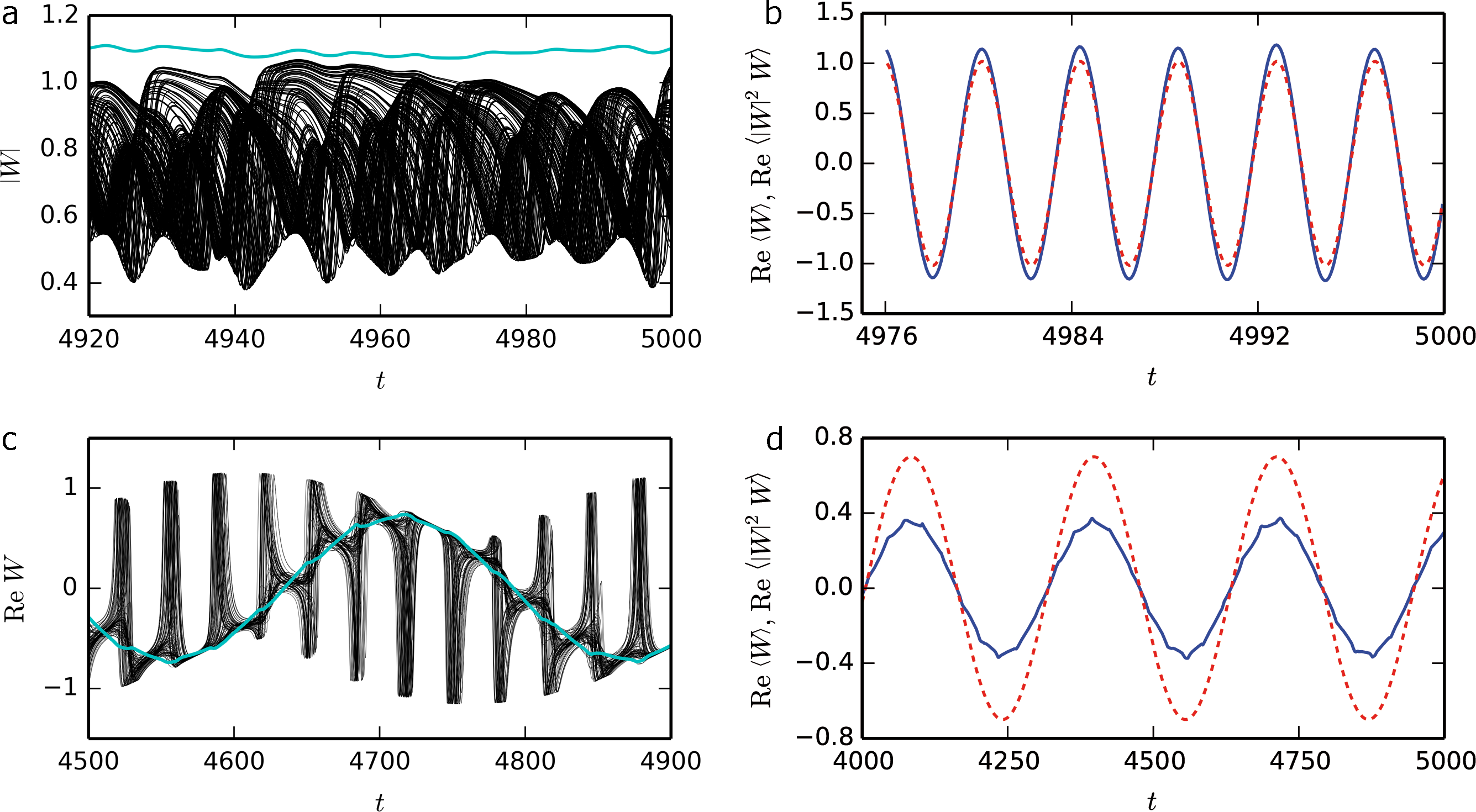}
  \caption{(Color online) (a,c) Type I and II chimera states, modulus and real part
    versus time, respectively. The population splits 
    into two groups, one being synchronized (cyan, gray) and one being
    desynchronized (black). (b,d) Linear average, Re $\left< W
    \right>$, (dashed lines) and nonlinear average, Re $\left< \left|
        W \right|^2 W \right>$, (solid lines) versus time for type I
    chimeras (b) and type II chimeras (d). }
  \label{fig:dynamics_averages}
\end{figure*}

The dynamics of the oscillator population, Eq.~\eqref{eq:SL_ensemble}, are determined by three
parameters, namely $c_2$, $\nu$ and $\eta$, where $\eta$ is
controlled via initial conditions and acts effectively as the 
coupling strength. We numerically solved
Eq.~\eqref{eq:SL_ensemble} using an implicit Adams method with
timestep $dt = 0.01$ for $N=1000$ oscillators and random initial
conditions fulfilling the conservation law in
Eq.~\eqref{eq:conservation_law}. The simulation results reveal two
types of clustering dynamics:
amplitude clusters as depicted in Fig.~\ref{fig:complexp}a and
modulated amplitude clusters as depicted in
Fig.~\ref{fig:complexp}f. 
In the amplitude cluster state the ensemble splits into two groups 
that oscillate with an amplitude difference and a small, fixed phase 
difference. The modulated amplitude cluster state can be described as 
an overall uniform oscillation that is modulated by an additional 
oscillation of the two groups around the mean value in antiphase, 
giving rise to quasiperiodic motion.
We studied these cluster solutions in
detail in Ref.~\cite{Schmidt_PRE_2014}, where we could show that the
amplitude clusters bifurcate off the synchronized solution
(Fig.~\ref{fig:complexp}c) via a pitchfork bifurcation. The modulated
amplitude clusters are created in a secondary Hopf bifurcation and the
two types of clusters are connected by a saddle-node of infinite
period (SNIPER) bifurcation. The cluster formation is the first
symmetry-breaking step rendering chimera states possible as it
produces first of all two different groups. Indeed, in the vicinity of
the two types of clusters we also observe two associated types of
chimera states, as shown in Figs.~\ref{fig:complexp}b and e,
respectively. The first type obviously inherited the property that the 
two groups are separated by an amplitude difference. Thus, starting 
from the amplitude cluster state, the group with the smaller radius got 
desynchronized. The second type of chimeras also shares properties 
with the modulated amplitude clusters, but this will be discussed 
below, where it becomes more apparent.
Type II chimeras (Fig.~\ref{fig:complexp}e) could actually be identified
with the chimera states found during the photoelectrodissolution of
n-type silicon \cite{Schmidt_Chaos_2014, Schmidt_arXiv_ECCII, Schoenleber_NJP_2014}. It bridges the
gap between the cluster solution in Fig.~\ref{fig:complexp}f and
completely irregular dynamics in Fig.~\ref{fig:complexp}d. In 
contrast, type I chimeras mediate between the
cluster solution in Fig.~\ref{fig:complexp}a and the synchronized
state in Fig.~\ref{fig:complexp}c.

To gain a better understanding of the temporal dynamics in the chimera
states, we depict $\left| W_k \right|$ and Re $W_k$ versus time for type I and II
chimeras in Figs.~\ref{fig:dynamics_averages}a and c, respectively.
The synchronized group is marked with cyan (gray) color and the incoherent
group is plotted in black. In the type I chimera there is a clear
separation of the groups by an amplitude difference. This state is in fact unstable, as we observe
heteroclinic transitions between the type I chimera and two other
cluster states on a large timescale. This will be discussed below. In contrast, the second type of chimeras seems to
be stable, as we could not observe a break down in the
simulations up to $T=1\cdot 10^6$. The incoherent oscillators in this type II chimera show
a nearly-periodic spiking behavior (which is not performed by all incoherent
oscillators at the same time). This is a property inherited from the
modulated amplitude clusters (Fig.~\ref{fig:complexp}f). The frequency
of the spiking is given by the frequency of the modulational
oscillations that are a result of a secondary Hopf bifurcation
\cite{Schmidt_PRE_2014}. The dynamics show that the separations
into incoherent and coherent groups occur via the clustering
mechanism, for both types of chimeras.

Type I chimeras can also be found with a linear global
coupling. Daido \& Nakanishi \cite{Daido_PRL_2006} and also Nakagawa
\& Kuramoto \cite{Nakagawa_PTP_1993} describe a state that seems to
be such a chimera state, but they do not
identify them as such. Only later they have been identified as chimera
states \cite{Sethia_PRL_2014}. In fact, the nonlinear global coupling we consider behaves effectively 
like a linear global coupling in case of type I dynamics. This is 
visualized in Fig.~\ref{fig:dynamics_averages}b, where we plot the 
linear part of the coupling $\left< W \right>$ as a red dashed line and 
the nonlinear part $\left< \left| W \right|^2 W \right>$ as a blue solid line. 
We see that the nonlinear term is also sinusoidal, i.e. $\left< \left| W \right|^2 W \right> \propto \left< W 
\right>$, yielding an effective overall linear behaviour of the 
coupling. Since this implies $\left< \left| W_k \right|^2 W_k \right> = 
\left< r_k^3 e^{i\phi_k} \right> \propto \left< r_k e^{i\phi_k}
\right>$, averaging leads to vanishing nonlinear effects in the global
coupling for type I chimeras.
In contrast, in case of type II chimeras the dynamics of $\left< \left| W \right|^2 W \right>$ is highly nonlinear, as shown in 
Fig.~\ref{fig:dynamics_averages}d. We conclude that type II 
dynamics might not be observable with a solely linear global coupling. 

\begin{figure}[h]
  \centering
  \includegraphics[width=8.6cm]{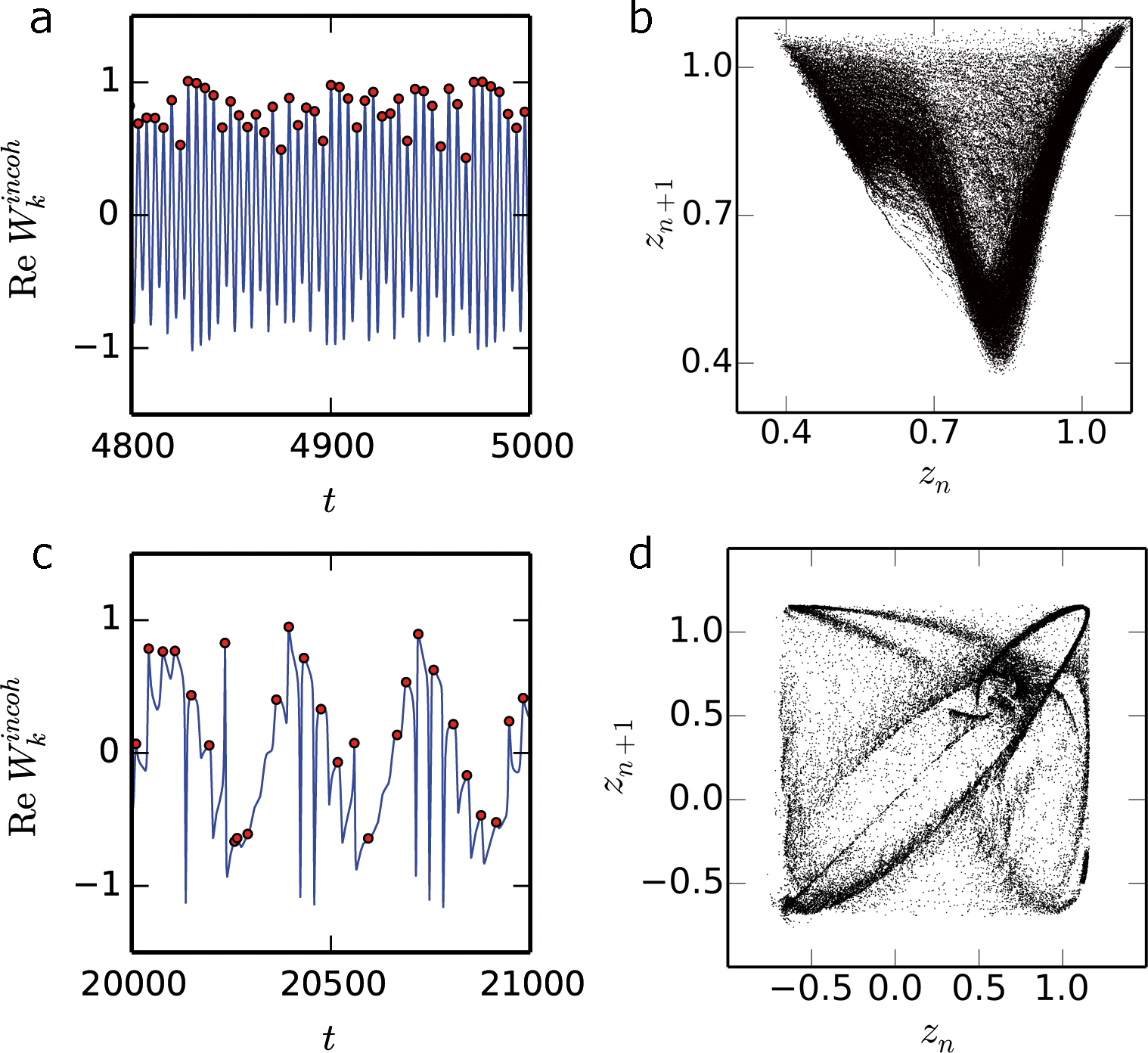}
  \caption{(Color online) Chaos in the chimera states. (a,c) Samples of timeseries of incoherent
    oscillators for type I and II chimeras, respectively. Identified peaks are marked with circles. (b,d) Next-maximum
    maps for the peaks in (a,c).}
  \label{fig:chaos}
\end{figure}

Furthermore, we looked at time series of individual oscillators in the incoherent 
groups. Examples are depicted in Figs.~\ref{fig:chaos}a and c for type 
I and II chimeras, respectively. As a simple test for chaoticity, next-maximum maps for the 
timeseries are shown in Figs.~\ref{fig:chaos}b and d, respectively. 
Both next-maximum maps are highly nontrivial and structurally very 
different. We see this as a clear indication that the dynamics in the 
incoherent parts of the two types of chimeras take place on
different types of chaotic attractors. In the case of type II
chimeras, the incoherent dynamics inherits properties from the motion
on the torus existing at close-by parameter values, while no torus
exists in the neighborhood of type I chimeras.

\begin{figure}[h]
  \centering
  \includegraphics[width=8.6cm]{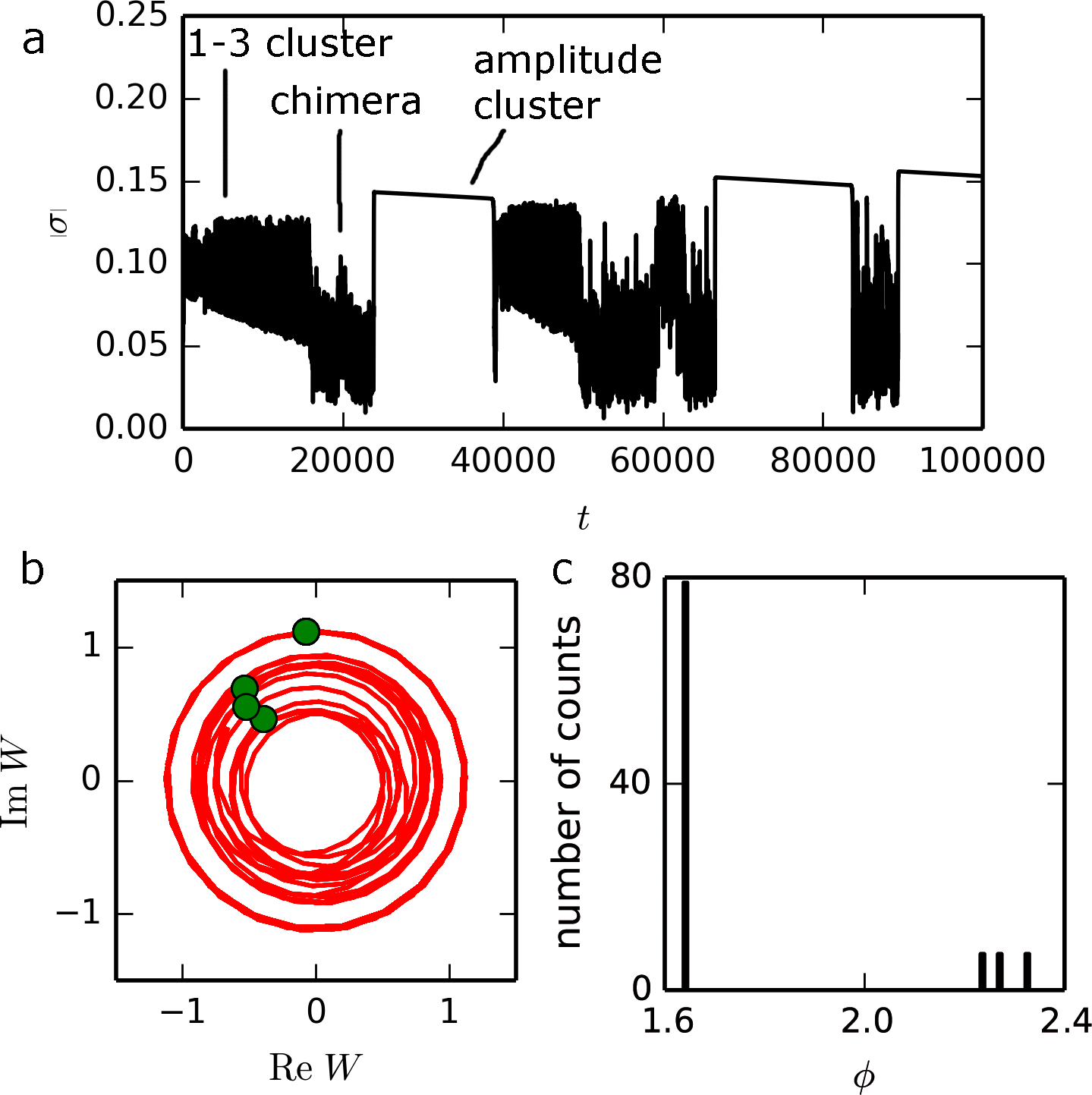}
  \caption{(Color online) Heteroclinic connections between type I chimeras, 1-3 
    cluster states and amplitude clusters for $N=100$ oscillators. (a) Trajectory of $\left| 
    \sigma \right|$ in time showing the transitions between the 
    different states. (b) Exemplary dynamics of the 1-3 cluster state 
    in the complex plane: lines depict time evolution and dots 
    represent the configuration of the oscillators at one timestep. 
    (c) Histogram of phases in the 1-3 clusters state showing that it 
    consists of 1 large cluster and 3 small clusters of approximately 
    the same size.}
  \label{fig:heteroclinic}
\end{figure}

As already mentioned, type I chimeras are unstable and we observe 
heteroclinic connections. To visualize this we define a measure 
characterizing the different dynamical states. The natural choice of 
the Kuramoto order parameter is inappropriate here, because of the 
strong amplitude fluctuations and since $\left< W \right> = \eta 
\exp \left( - i \nu t \right)$ at all times. Therefore we use the 
variance $\sigma = \left< W^2 \right> - \left< W \right>^2$. An 
exemplary timeseries of $\left| \sigma \right|$ for parameters of 
type I chimeras is shown in Fig.~\ref{fig:heteroclinic}a.

Three qualitatively different regimes can be identified and after an
initial transient, the system randomly settles first to one of them;
in the trajectory shown it is a 1-3 cluster state. The 
dynamics in this state are depicted in Fig.~\ref{fig:heteroclinic}b 
and the phase distribution at one timestep is shown in a histogram in 
Fig.~\ref{fig:heteroclinic}c. This state consists of one large cluster 
and three small clusters of approximately the same size. The measure 
$\left| \sigma \right|$ exhibits strong variations around a value of 
approximately 0.1. Then around t = 15000 the 1-3 cluster state breaks 
down and the system moves to a new state that exhibits fluctuations of 
$\left| \sigma \right|$ around 0.05: the type I chimera state. After 
approximately $\Delta t = 10000$ we observe another transition to a 
state with nearly constant $\left| \sigma \right|$. This is the 
amplitude cluster state as depicted in Fig.~\ref{fig:complexp}a. 
Figure~\ref{fig:heteroclinic}a suggests that transitions between these
three states follow in a non-cyclic and non-periodic sequence. Thus, though being reminiscent of a 
heteroclinic orbit, the dynamics possesses a further peculiar, 
unpredictable feature.

In summary, we found numerically two types of chimera states in the 
vicinity of two types of clusters. The chimera states inherit 
properties from the respective cluster states. We conclude that the 
clustering mechanism is a first symmetry-breaking step sufficient for 
chimera states to occur in oscillatory systems with uniform global coupling. 
It differentiates the system into two groups thereby rendering it
bistable. Oscillators in the two states respond effectively different to the coupling due
to nonlinear amplitude effects. Note that as a consequence, this mechanism will
not give rise to chimeras in ensembles of phase oscillators, where
other mechanisms may render their formation possible \cite{Yeldesbay_PRL_2014}.
Furthermore, we demonstrated that the chimera states can mediate between cluster 
states and completely incoherent behavior as well as between cluster
states and synchrony. This leads us to the conclusion that chimera
states might appear spontaneously in many globally coupled systems, as a clustering mechanism and the possibility of 
amplitude variations are sufficient features a system has to exhibit.

\begin{acknowledgments}
We thank Vladimir Garc\'{i}a-Morales and Konrad Sch\"{o}nleber for 
fruitful discussions. Furthermore, we gratefully acknowledge financial 
support from the \textit{Deutsche Forschungsgemeinschaft} (Grant no. 
KR1189/12-1), the \textit{Institute for Advanced Study, Technische 
Universit\"{a}t M\"{u}nchen}, funded by the German Excellence 
Initiative and the cluster of excellence \textit{Nanosystems 
Initiative Munich (NIM)}.
\end{acknowledgments}

\bibliography{lit}

\end{document}